\documentclass[aps,prb,twocolumn,floatfix,superscriptaddress]{revtex4}

\usepackage{amsmath}
\usepackage{amssymb}
\usepackage{times}
\usepackage{braket}
\usepackage[pdftex]{graphicx}
\usepackage{blindtext}
\usepackage{color}

\renewcommand{\vec}[1]{\boldsymbol{#1}}

\newcommand{\change}[1]{\textcolor{black}{#1}}
 
\begin{document}

\title{Skyrmion ratchet propagation: Utilizing the skyrmion Hall effect in AC racetrack storage devices}

\author{B{\"o}rge G{\"o}bel}
\email[Corresponding author. ]{boerge.goebel@physik.uni-halle.de}
\affiliation{Institut f\"ur Physik, Martin-Luther-Universit\"at Halle-Wittenberg, D-06099 Halle (Saale), Germany}
%\affiliation{Max-Planck-Institut f\"ur Mikrostrukturphysik, D-06120 Halle (Saale), Germany}

\author{Ingrid Mertig}
\affiliation{Institut f\"ur Physik, Martin-Luther-Universit\"at Halle-Wittenberg, D-06099 Halle (Saale), Germany}

\date{\today}

\begin{abstract}
\noindent \textbf{Abstract.} Magnetic skyrmions are whirl-like nano-objects with topological protection. When driven by direct currents (DC), skyrmions move but experience a transverse deflection. This so-called skyrmion Hall effect is often regarded a drawback for memory applications. Herein, we show that this unique effect can also be favorable for spintronic applications: We show that in a racetrack with a broken inversion symmetry, the skyrmion Hall effect allows to translate an alternating current (AC) into a directed motion along the track, like in a ratchet. We analyze several modes of the ratchet mechanism and show that it is unique for topological magnetic whirls. We elaborate on the fundamental differences compared to the motion of topologically trivial magnetic objects, as well as classical particles driven by periodic forces. Depending on the exact racetrack geometry, the ratchet mechanism can be soft or strict. In the latter case, the skyrmion propagates close to the efficiency maximum.
\end{abstract}

%\pacs{aaa}

\maketitle
\noindent\textbf{Introduction}\\
When spatial symmetries in a physical system are broken, periodic or even randomly oriented perturbations can lead to a net force and a directed propagation \cite{feynman1966feynman}. For example, a charged particle in a periodic but asymmetric environment, can be driven by an alternating electric field; it will move step-by-step from one energy minimum to the next minimum along the direction with the smaller potential gradient. This so-called ratchet mechanism has been widely explored, especially for using thermal fluctuations to drive molecular motors \cite{reimann2002brownian,hanggi2009artificial} or even in spintronics to unidirectionally move magnetic solitons in a shift register \cite{lavrijsen2013magnetic} or to drive spins with alternating voltages \cite{flatte2008one,costache2010experimental}.

Besides analyzing individual spins, over the recent decade, the applicability of non-collinear spin textures in spintronics has been explored intensively.  
For example, the ratchet propulsion \cite{himeno2008domain,franken2012shift, piao2011ratchet} and diode behavior of magnetic domain walls have been investigated \cite{whyte2015diode}. Herein, we consider fundamentally different spin textures: the nano-sized magnetic skyrmions~\cite{bogdanov1989thermodynamically,muhlbauer2009skyrmion,yu2010real} which are whirl-like magnetic objects that possess particle-like properties. Their integer topological charge~\cite{nagaosa2013topological}
\begin{align}
N_\mathrm{Sk}=\frac{1}{4\pi}\int\vec{m}(\vec{r})\cdot\left(\frac{\partial\vec{m}(\vec{r})}{\partial x}\times \frac{\partial\vec{m}(\vec{r})}{\partial y}\right)\,\mathrm{d}^2r=\pm 1\notag
\end{align}
brings about a topologically enforced stability of skyrmions. Even under deformation, their unique properties remain. 

For this reason, skyrmions are often considered to become bits in racetrack data storage devices in the future~\cite{parkin2004shiftable,parkin2008magnetic, parkin2015memory,sampaio2013nucleation, fert2013skyrmions,yu2017room}. This concept is based on the controlled generation and electrical motion of bits in thin nano-stripes; the presence versus the absence of magnetic skyrmions at predefined positions corresponds to `1' and `0' bits. Since the racetrack operates purely electrically and since skyrmions can be driven by very small current densities \cite{jonietz2010spin}, the racetrack is advantageous compared to typical RAM or HDD memories~\cite{parkin2008magnetic}. However, due to the topological charge of the skyrmions, the objects do not move parallel to the applied current but are deflected towards the edge of the racetrack~\cite{nagaosa2013topological,zang2011dynamics, iwasaki2013current,jiang2017direct,litzius2017skyrmion, tomasello2014strategy}. 

An enormous effort has been undertaken to find solutions for suppressing this so called skyrmion Hall effect, for example modifying the spin torque~\cite{zhang2015skyrmion,gobel2018magnetic} or using alternative magnetic quasiparticles~\cite{gobel2020beyond} with a vanishing topological charge (like antiferromagnetic skyrmions~\cite{barker2016static,zhang2016magnetic,zhang2016antiferromagnetic, gobel2017afmskx,legrand2020room,dohi2019formation} or skyrmioniums~\cite{zhang2016control,goebel2019electrical,zhang2018real}) or nano-objects with a broken rotational symmetry (like antiskyrmions~\cite{nayak2017magnetic,jena2020elliptical} or bimerons~\cite{gobel2018magnetic,kharkov2017bound,gao2019creation}). 
%Today, these solutions mostly remain theoretical suggestions. 
However, as we show in the present paper, the skyrmion Hall effect may even be a favorable feature that can be utilized instead of trying to suppress it.

\begin{figure}[b!]
  \centering
  \includegraphics[width=\columnwidth]{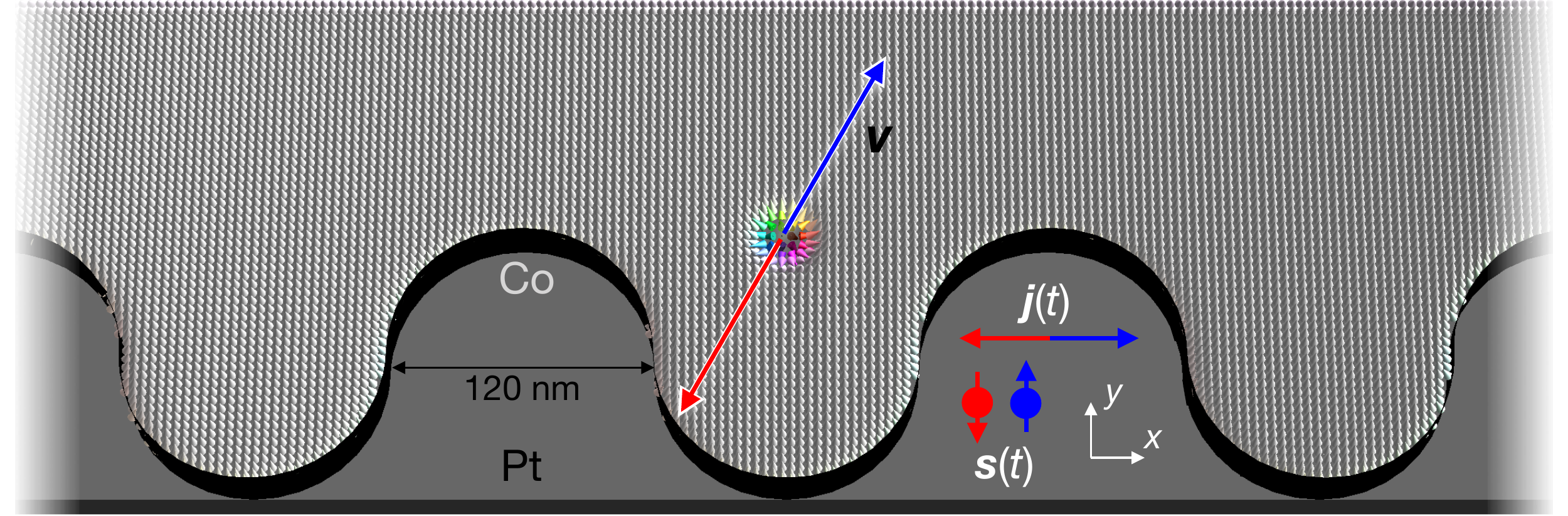}
  \caption{\textbf{AC propulsion of skyrmions in asymmetric racetracks.} A Co layer (small arrows indicate magnetic moments whose color corresponds to their orientation; white $+z$, black $-z$) is interfaced with a Pt layer (dark gray) so that an applied current $\vec{j}(t)\parallel \pm \vec{x}$ leads to an injection of spins $\vec{s}(t)\parallel \pm \vec{y}$ from the perpendicular direction. In dependence of the orientation of $\vec{s}$, the skyrmion moves along $\pm\vec{v}$ (red and blue). Due to the different interaction with the two edges, the skyrmion can experience a net propulsion along the positive $x$ direction.}
  \label{fig:overview}
\end{figure}

We simulate a skyrmion ratchet that is driven by temporally periodic  electric currents. By breaking the mirror symmetry of the racetrack along the racetrack's width $y$ (i.\,e. a broken mirror axis along the racetrack direction $x$), the skyrmion Hall effect translates the alternating current into a directed motion along the track ($x$ direction, as indicated in Fig. \ref{fig:overview}). While in symmetric racetracks the skyrmion can only move when direct currents (DC) are applied, the modified setup allows to operate a racetrack with alternating currents (AC). As we will show, this ratchet mechanism is strongly tied to the non-trivial topology of the skyrmions.

Using the inate topological properties of skyrmions is in contrast to other predicted skyrmion ratchet devices that rely on asymmetric anisotropy gradients \cite{wang2018efficient} or pinning sites \cite{ma2017reversible} along the racetrack, or on using skyrmion-shape modulations to generate a weak net forces from periodic magnetic fields \cite{moon2016skyrmion,chen2019skyrmion}. Instead, the here presented approach, based on the topology of skyrmions, is more related to the observed directed rotation of a skyrmion lattice due to thermal fluctuations of the chiral spin texture \cite{mochizuki2014thermally} and to the recently predicted skyrmion diode \cite{zhao2020ferromagnetic} that allows for skyrmion propagation only along one direction when a current is applied. \change{The results are related to the Magnus-induced dynamics of skyrmions and other particles as reported in Refs. \cite{reichhardt2015magnus,reichhardt2020dynamics}. However, since here a confined geometry is used instead of periodic boundary conditions, the role of the asymmetric confinement becomes relevant for translating an AC current into a directed propulsion (anti)parallel to it.} \\
\\
\textbf{Results}\\
\textbf{Skyrmion motion in symmetric racetracks.}
When physical objects are driven by periodic external forces, they will exhibit an oscillating motion. A net propulsion is only possible when the motion is not equivalent along the two opposite directions, e.\,g. due to friction or an asymmetric potential. The same applies to the motion of magnetic spin textures under the influence of spin torques. The motion of a magnetic skyrmion under spin-orbit torques is simulated by propagating the Landau-Lifshitz-Gilbert equation \cite{landau1935theory,gilbert1955lagrangian,slonczewski1996current} 
 of every magnetization cell (see Methods section for details). We use this method throughout this paper and present micromagnetic simulations for which we have employed mumax3 \cite{vansteenkiste2011mumax, vansteenkiste2014design}. To establish a better understanding of the results of these simulations and the underlying mechanisms, the motion of a skyrmion will also be effectively described by the Thiele equation \cite{thiele1973steady,gobel2018overcoming}
\begin{align}
-4\pi N_\mathrm{Sk}b\,\vec{e}_z\times \dot{\vec{r}}-\underline{D}\alpha b\,\dot{\vec{r}}-Bj\underline{I}\,\vec{s}=\nabla U(\vec{r}). \label{eq:thiele}
\end{align}
This description condenses the nano-object to a single point positioned at $\vec{r}$. The non-collinearities of a skyrmion are covered by the topological charge $N_\mathrm{Sk}$, the dissipative tensor $D_{ij}=\int\partial_{x_i}\vec{m}(\vec{r})\cdot\partial_{x_j}\vec{m}(\vec{r})\,\mathrm{d}^2r$ and the torque tensor $I_{ij}=\int[\partial_{x_i}\vec{m}(\vec{r})\times\vec{m}(\vec{r})]_{x_j}\,\mathrm{d}^2r$ \change{that only have to be calculated once, as long as the skyrmion does not significantly changes its shape during propagation}. The first term of the Thiele equation \eqref{eq:thiele} describes a transverse motion due to the topological charge $N_\mathrm{Sk}$. The second term describes dissipation, since the dissipative tensor $\underline{D}$ is typically diagonal. The coefficients are: Gilbert damping $\alpha$ and \change{$b=M_sd_z/\gamma_e$}, where $M_s$ is the saturation magnetization, $d_z$ the thickness of the magnetic layer and $\gamma_e$ the gyromagnetic ratio.

The spin-orbit torque enters the third term. If one considers an interface of a ferromagnet and a heavy metal (like Co/Pt, as we will consider in the following; cf. Fig. \ref{fig:overview}), an applied current (density $\vec{j}$) flows mainly in the heavy metal. Here, the spin Hall effect leads to the injection of spins $\vec{s}$ into the ferromagnet from the perpendicular direction. These spins exhibit a torque onto the magnetic moments that constitute a magnetic skyrmion. The moments reorient collectively and the center coordinate of the skyrmion $\vec{r}$ can change. In the coordinate system that is defined in Fig. \ref{fig:overview} we assume a current flowing along the racetrack $\vec{j}\parallel\pm\vec{e}_x$ that leads to the injection of spins $\vec{s}\parallel\mp\vec{e}_y$ (\change{dimension-less quantity;} the sign depends on the sign of the spin Hall conductivity). The efficiency $B=\frac{\hbar}{2e}\Theta_\mathrm{SH}$ of the spin-orbit torque depends on the spin Hall angle $\Theta_\mathrm{SH}$.

The right side of the Thiele equation \eqref{eq:thiele} describes the interaction of the nano-object with other \change{non-collinear spin textures, defects (attractive or repulsive)\cite{fernandes2018universality,muller2015capturing,muller2016edge, hanneken2016pinning, castell2019accelerating},} or with the confinement \change{\cite{zhang2015skyrmion,chen2017skyrmion}}. \change{Here, we consider no defects and no other spin textures but only the repulsive interaction with the confinement.} In a symmetric racetrack, the potential is symmetric along the racetrack width $U(y)=U(-y)$ giving rise to antisymmetric forces $\vec{F}_\mathrm{edge}=-\nabla U(\vec{r})$ along the $\pm y$ direction. This means the confinement always pushes the object to the center.  Furthermore, this symmetric potential does not \change{directly} generate forces along the racetrack direction $x$. \change{Note, however, that it still affects $\dot{\vec{r}}$ and therefore also the orientation of the gyroscopic force. This indirectly leads to an acceleration of the motion along the racetrack direction when a skyrmion slides along the edge (cf. Supplementary Fig. 1). However, this effect is symmetric:} When a spin texture, topologically trivial or not, is driven by a periodic current in this environment, the net motion vanishes.

\begin{figure}[t!]
  \centering
  \includegraphics[width=\columnwidth]{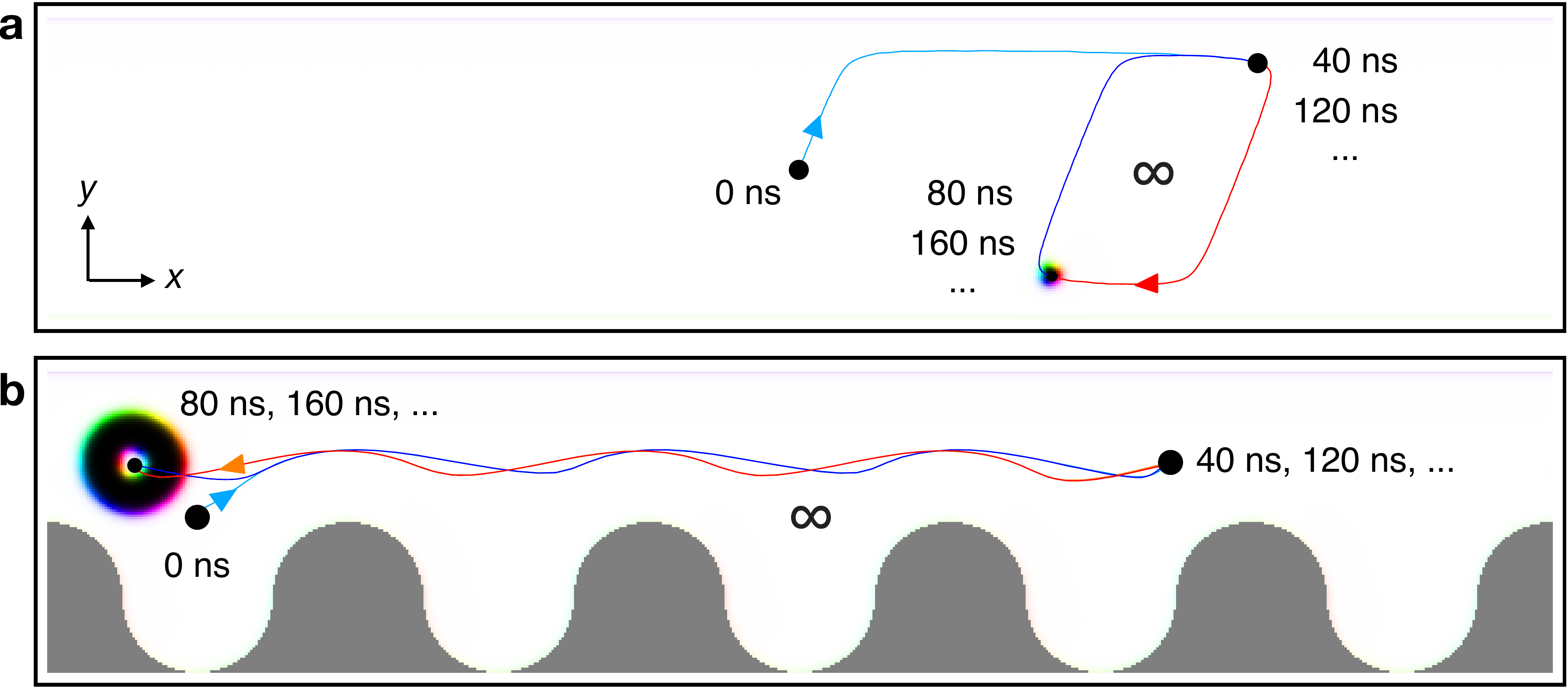}
  \caption{\textbf{Lack of net propulsion in symmetric systems or for topologically trivial nano-objects.} \textbf{a}, Motion of a skyrmion under the influence of an alternating current in a symmetric racetrack. The current is characterized by $T=80\,\mathrm{ns}$ and $j_{x,\mathrm{max}}\Theta_\mathrm{SH}=3.0\,\frac{\mathrm{MA}}{\mathrm{cm}^2}$. After an initial half-period $T/2 = 40\,\mathrm{ns}$ (cyan trajectory), the skyrmion periodically moves on a closed trajectory without a net propagation (red and blue). The racetrack dimensions are $1200\,\mathrm{nm}\times 240\,\mathrm{nm}\times 2\,\mathrm{nm}$ with periodic boundary conditions along the track direction. \textbf{b}, Motion of a skyrmionium in a modified racetrack. The pockets have been implemented as was presented in Fig. \ref{fig:overview}. The skyrmionium moves periodically between several pockets of the racetrack without a skyrmion Hall effect and without a net propagation. The setup is a Co layer on Pt with parameters given in the Methods section. Animated versions of these two panels are shown in the Supplementary Movies 1 and 2, respectively.}
  \label{fig:trivial}
\end{figure}

We visualized this scenario for the example of a N\'{e}el skyrmion in a Co/Pt sample in Fig \ref{fig:trivial}a. The symmetry of this texture (only $D\equiv D_{xx}=D_{yy}\neq0$ and $I\equiv I_{xy}=-I_{yx}\neq0$) simplifies the Thiele equation to
\begin{align}
4\pi N_\mathrm{Sk}b
\begin{pmatrix}
-v_y\\
v_x
\end{pmatrix}
+D\alpha b
\begin{pmatrix}
v_x\\
v_y
\end{pmatrix}
+BIj
\,\vec{e}_x
=\vec{F}_\mathrm{edge},\notag
\end{align}
with $\vec{F}_\mathrm{edge}=F(y)\vec{e}_y$, due to the symmetric racetrack. Since the magnetic background is magnetized along the positive $z$ direction, $N_\mathrm{Sk}=-1$. The motion has been simulated using micromagnetic simulations in Fig. \ref{fig:trivial}a. For details and the simulation parameters of the Co/Pt bilayer we refer to the Methods section. 

In this simulation and throughout the paper, we consider a periodic current $\vec{j}=j_{x,\mathrm{max}}\sin(\frac{2\pi}{T}t)\vec{e}_x$.
%, leading to $\vec{s}=-s_0\sin(\frac{2\pi}{T}t)\vec{e}_y$.
Starting from $t=0\,\mathrm{ns}$, a positive current is applied. Consequently, the skyrmion is pushed towards the upper edge (positive $x$ and $y$ direction of motion). This partially transverse motion is the signature of the skyrmion Hall effect and is caused by the gyroscopic force: the first term in the Thiele equation \eqref{eq:thiele}. The trajectory of the skyrmion follows the skyrmion Hall angle $\theta_\mathrm{sk}=\arctan \frac{-4\pi N_\mathrm{Sk}}{\alpha D}\approx 70.0^\circ$, numerically determined from $D=15.3$ for the used simulation parameters. Once the skyrmion reaches the edge, it can survive and glide along the edge of the racetrack ($x$ direction). The edge potential $U$ contributes a force $\vec{F}_\mathrm{edge}$ along $-y$, to the middle of the racetrack, so that the gyroscopic force is compensated. Note that there exists a critical current density beyond which the gyroscopic force becomes too large so that the skyrmion annihilates at the edge. \change{Furthermore, for current densities close to $j_{x,\mathrm{max}}$, the skyrmion deforms due to the fast propagation. As a consequence, the tensors $\underline{D}$ and $\underline{I}$ change which affects $\theta_\mathrm{sk}$ slightly, as was analyzed in Ref. \cite{gobel2018overcoming}. Here, the current density $j_{x,\mathrm{max}}$ is sufficiently small, so that the skyrmion remains stable and does not significantly deform.}

After half a period $T/2$, the transient process has been completed and the skyrmion starts to oscillate in a closed loop (Fig. \ref{fig:trivial}a): First, the current becomes negative, so the skyrmion moves along the negative $-x$ and $-y$ directions towards the lower edge where it glides along the $-x$ direction. After a full period $T$, the current becomes positive again, so that the skyrmion starts to move back to the upper edge of the racetrack. On average, no net motion has been achieved. This is true for all period durations $T$ and all reasonable current densities $j_{x,\mathrm{max}}$.\\
\\
\textbf{Skyrmion ratchet mechanism in asymmetric racetracks.}
To achieve a net motion, the \change{racetrack's symmetry has to be broken. Here, the confinement potential $U(\vec{r})$ \cite{iwasaki2014colossal,navau2016analytical,song2017skyrmion,muller2017magnetic} plays an important role, because it can introduce} $\vec{F}_\mathrm{edge}$ components along the racetrack. The racetrack geometry has to be modified. However, since the racetrack has to extend (in principle) infinitely along $x$, the geometry manipulation can only be periodic. In Fig. \ref{fig:overview} the considered geometry is presented: One edge of the racetrack remains straight while the other is periodically modulated. Depending on the position of the nano-object, it experiences a force component along $+x$ or $-x$.

However, for topologically trivial objects, $N_\mathrm{Sk}=0$, like the skyrmionium~\cite{zhang2016control,goebel2019electrical,zhang2018real} (one skyrmion inside of an oppositely magnetized other skyrmion), this does not generate a net driving force because the modified potential only periodically decelerated and accelerates the object: The skyrmionium experiences hardly any transverse deflection and the decelerating and accelerating forces from the boundaries cancel on average over time. As a consequence, the skyrmionium merely oscillates periodically, cf. Fig. \ref{fig:trivial}b.

\begin{figure}[t!]
  \centering
  \includegraphics[width=\columnwidth]{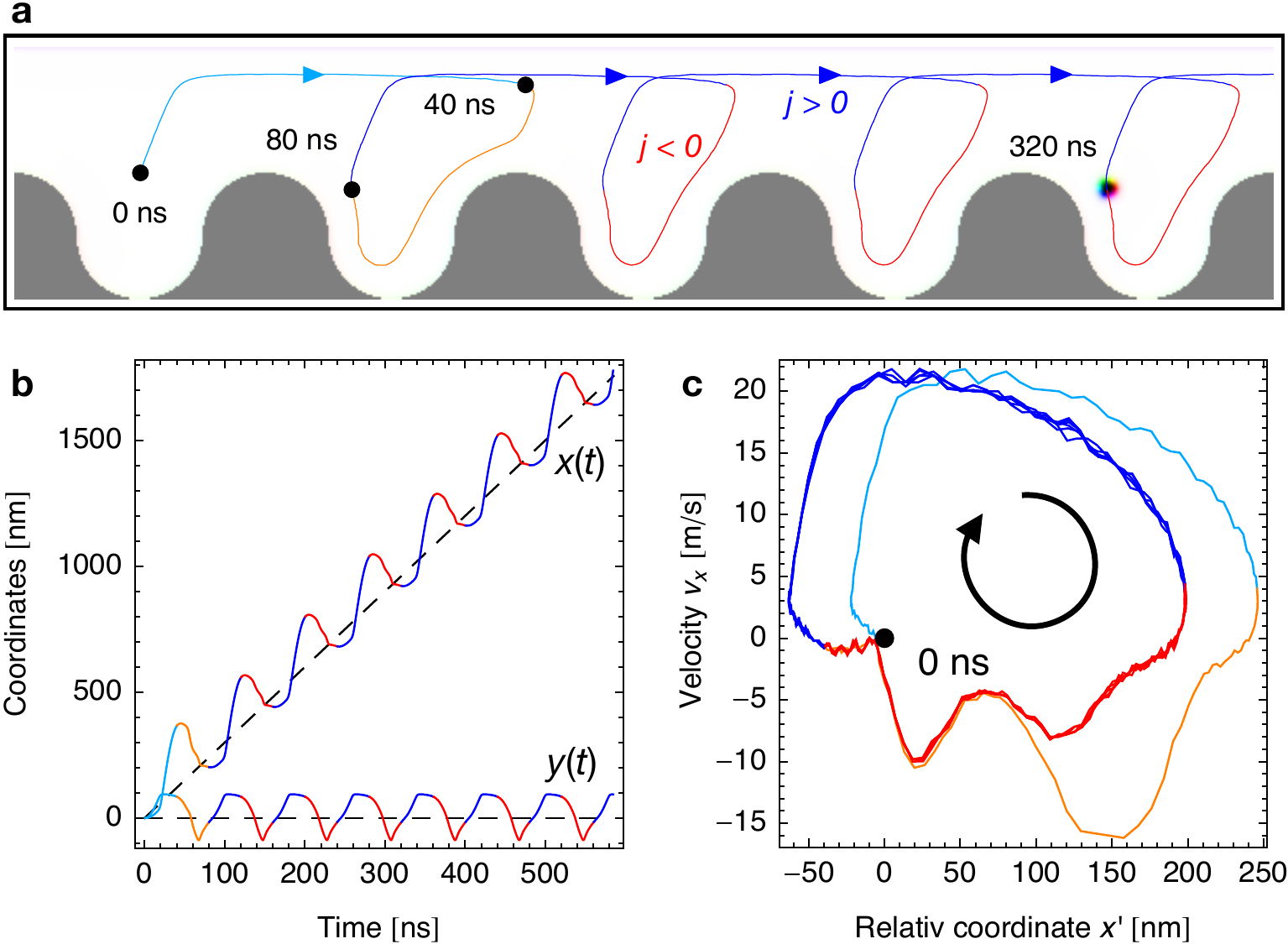}
  \caption{\textbf{Skyrmion propulsion by alternating currents via a ratchet mechanism.} \textbf{a}, Due to the asymmetric racetrack design and the transverse deflection due to the skyrmion's topological charge, the skyrmion moves differing distances for positive (blue) and negative (red) current densities. The trajectory shows that after the first transient period $T$ (cyan and orange), on average, the skyrmion has moved one cell distance $L$ (blue and red) per period $T$. \textbf{b}, The $x$ and $y$ coordinates over time. In both cases the curves oscillate around a linear function with slope $\overline{v}_x=L/T$ and $\overline{v}_y=0$, respectively. \textbf{c}, The $x'=x-\overline{v}_xt$ dependence of $v_x$ exhibits several closed loops indicating a quasi-periodic motion. The diagram describes the motion of the skyrmion with respect to the average motion. Here, the current is characterized by $j_{x,\mathrm{max}}\Theta_\mathrm{SH}=3.0\,\frac{\mathrm{MA}}{\mathrm{cm}^2}$, $T=80\,\mathrm{ns}$ and the distance between two neighboring cells is $L=240\,\mathrm{nm}$. An animated version of \textbf{a} is shown in Supplementary Movie 3.}
  \label{fig:ratchet}
\end{figure}

However, when we consider a skyrmion $N_\mathrm{Sk}=-1$ in the modified geometry, a net motion is observable, as is presented in Fig. \ref{fig:ratchet}. After the first transient period $T=80\,\mathrm{ns}$ of the alternating current (cyan and orange trajectories), the skyrmion effectively moves one spatial period $L=240\,\mathrm{nm}$ per current period (Fig. \ref{fig:ratchet}a) and oscillates around the average motion coordinate on a quasi-periodic trajectory (blue and red in Fig. \ref{fig:ratchet}b). The average velocity is determined by the geometry and the current period $\overline{\vec{v}}=L/T\,\vec{e}_x=3.0\frac{\mathrm{m}}{\mathrm{s}}\,\vec{e}_x$ (slope of the dashed line in Fig. \ref{fig:ratchet}b).

The quasi-periodic motion can be characterized by the $x'$-$v_x$ diagram in Fig. \ref{fig:ratchet}c, where $x'=x-\overline{v}_xt$ is the relative $x$ coordinate with respect to the coordinate of the average motion. After the first transient period (cyan and orange in Fig. \ref{fig:ratchet}c), the curve exhibits several almost identical closed loops (blue and red). The skyrmion moves on a periodic trajectory with respect to the position corresponding to the average motion. 

 We start our detailed discussion of the propulsion mechanism at $t=80\,\mathrm{ns}$ which is at the end of the first period $T$. This is because the first period constitutes a transient process because we start from a skyrmion in the center of its respective pocket and end up with a skyrmion that is positioned at the left side of the next pocket. 
 
The motion during the first half of the period $T/2$ is equivalent to the motion of the skyrmion in the symmetric racetrack; cf. Fig. \ref{fig:trivial}a. The skyrmion moves towards the upper edge and glides along the positive $x$ direction. The current parameters have been chosen such that the skyrmion has moved about $1.5\,L$, so that it is now positioned a the very right of the next racetrack pocket. For the next half period $T/2$, the current becomes negative and the skyrmion moves towards the lower edge where $\vec{F}_\mathrm{edge}$ depends strongly on the $x$ coordinate and even changes its orientation. The skyrmion moves along the whole modulated bottom edge of this pocket until the end of the first period is reached. The skyrmion ends up at the left of that pocket. Over the course of one period it has moved one pocket to the right. 

The key ingredient of any ratchet mechanism is to break the balance between forwards and backwards propagation. Here, the edge mediated force $\vec{F}_\mathrm{edge}$ accelerates the forwards motion but decelerates most parts of the backwards motion. This can be comprehended in detail in Fig. \ref{fig:ratchet}c.\\
\\
\textbf{Different propagation modes.}
The presented propulsion is parameter dependent and has been optimized to achieve a well-periodic motion. In the following, we investigate the influence of the current period $T$ and the amplitude $j_{x,\mathrm{max}}$ and present several propagation modes. 

\begin{figure}[t!]
  \centering
  \includegraphics[width=\columnwidth]{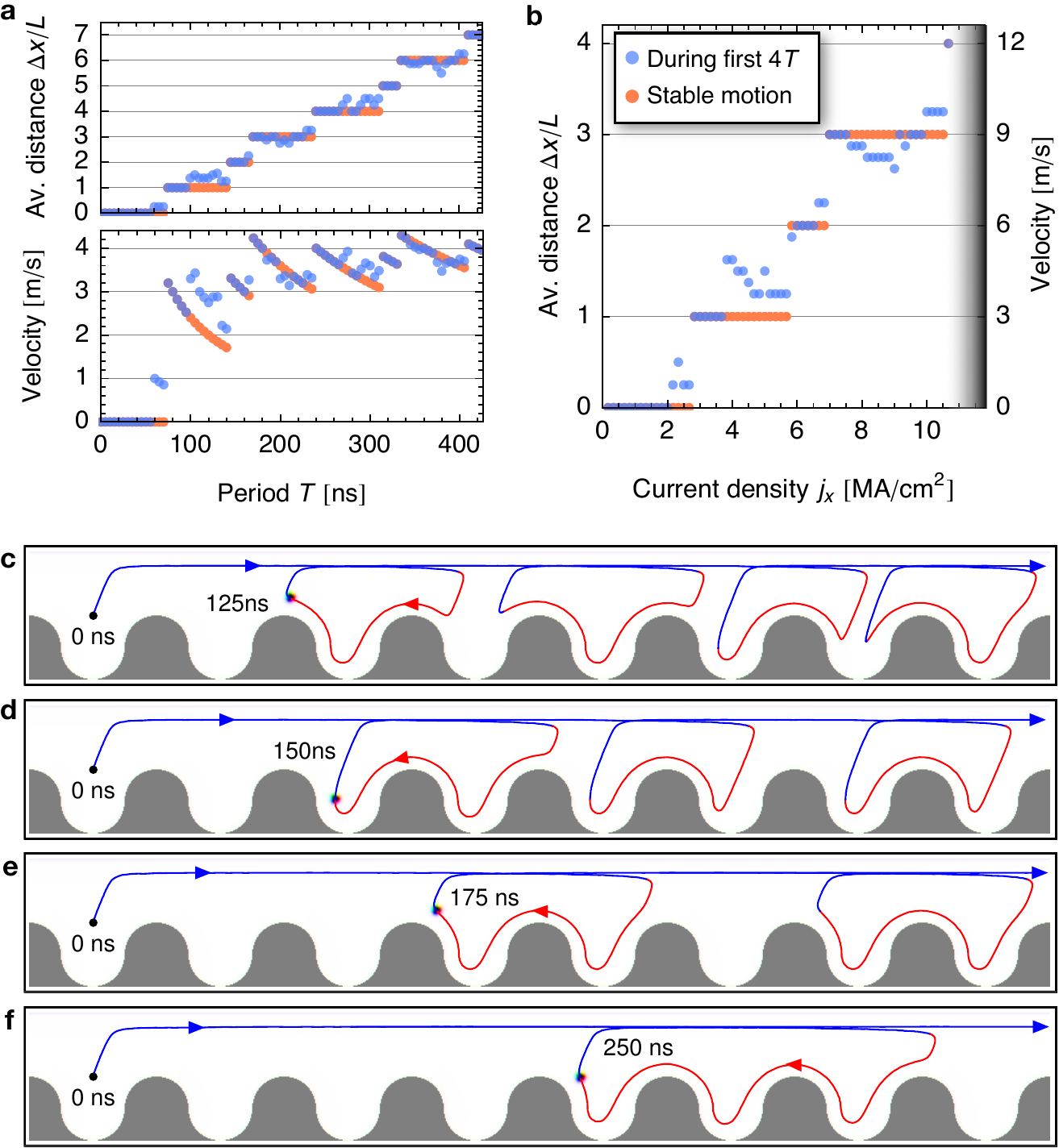}
  \caption{\textbf{Current amplitude and frequency dependence of the propagation mode.} \textbf{a}, Period dependence of the propagation distance. The top panel shows the \change{average travelled distance per period. For determining the blue data points, the first 4 periods of simulation time have been considered. The red data points have been determined after waiting until a steady propagation mode has been reached. This especially excludes the initial transient processes during propagation.} $L=240\,\mathrm{nm}$ is the cell size. The bottom panel shows the velocity; the same data but normalized by the \change{period $T$.} \textbf{b}, Current amplitude dependence of the \change{average} propagation distance for a fixed $T=80\,\mathrm{ns}$. \change{The colored data points correspond to the propagation including (blue) and excluding (red) the initial transient processes, as explained before.} In the gray shaded region, the skyrmion becomes unstable and annihilates at the racetrack's edge. \textbf{c}, Non-periodic motion between 1 and 2 steps per period for $T=125\,\mathrm{ns}$. The colored line indicates the skyrmion trajectory. \textbf{d}, Two-step propagation mode for $T=150\,\mathrm{ns}$. \textbf{e}, Three-step mode for $T=175\,\mathrm{ns}$. \textbf{f}, Four-step mode for $T=250\,\mathrm{ns}$. The current density is $j_{x,\mathrm{max}}\Theta_\mathrm{SH}=3.0\,\mathrm{MA}/\mathrm{cm}^2$ in \textbf{a,c-f}. Animated versions of panels \textbf{c-f} are shown in the Supplementary Movies 4-7, respectively.}
  \label{fig:modes}
\end{figure}

We start with the $T$ dependence when the current is kept constant $j_{x,\mathrm{max}}\Theta_\mathrm{SH}=3.0\,\mathrm{MA}/\mathrm{cm}^2$; see Fig. \ref{fig:modes}a. We simulate the skyrmion motion for the time span of $4T$ and determine the traveled distance. \change{This results in the blue data points for the average traveled distance $\Delta x$. For the determination of the red data points we have waited until a steady motion was reached to exclude the effect of transient processes that will be mentioned in the following.}

When the period $T$ is too small, no net propulsion is achieved. The skyrmion only oscillates in the starting pocket. For periods starting from $60\,\mathrm{ns}$, one finds that the initial period leads to a single skyrmion step. However, thereafter, the skyrmion merely oscillates in that pocket. This is a signature of the transient process in the beginning of the simulation: The first period behaves differently compared to the following periods because of inequivalent starting positions.

Starting from $T=75\,\mathrm{ns}$, the skyrmion always travels one step per period, as was extensively analyzed in Fig. \ref{fig:ratchet}. 
For larger periods, even multiple steps per periods can be achieved. In Fig. \ref{fig:modes}c a two-step propulsion mode is shown for $T=150\,\mathrm{ns}$. It becomes apparent that the skyrmion actually travels three steps forwards and one step backwards indicating a soft-ratchet behavior. For periods in between, one finds again signatures of the transient dynamics and non-periodic behavior; Fig. \ref{fig:modes}c: During the first periods, the skyrmion travels $2L$ and thereafter it only travels $1L$ per period. \change{This results in different values for the blue and red data points in Fig. \ref{fig:modes}a, including and excluding the initial transient process, respectively.} Generally speaking, by increasing the period, the traveled distance increases per period: In Figs. \ref{fig:modes}e,f three- and four-step propagation modes are presented for the periods $T=175\,\mathrm{ns}$ and $T=250\,\mathrm{ns}$, respectively.  For the latter mode, the skyrmion actually travels 6 steps forwards and 2 steps backwards, again showing that the mechanism corresponds to a soft ratchet.

As mentioned above, in the top diagram in Fig. \ref{fig:modes}a the \change{average skyrmion propulsion distance} is shown. It roughly shows a linear dependence but with several plateaus. In the bottom diagram the velocity is visualized, i.\,e. the distance has been normalized by the \change{period $T$}. Especially at higher $T$, the curve is almost constant meaning that the period does not allow to tune the average velocity significantly but it allows to tune the mode of propagation, i.\,e., the number of steps per period.

In Fig. \ref{fig:modes}b the current density dependence of the \change{average} traveled distance is presented for a fixed $T=80\,\mathrm{ns}$. Again, the dependence is roughly linear and several plateaus are present due to the ratchet nature of the propulsion mechanism. The curve increases up to about $10.5\,\mathrm{MA}/\mathrm{cm}^2$. Beyond this current density, the skyrmion becomes unstable and annihilates at the edge of the racetrack (gray area). In contrast to the $T$ dependence in panel a, here, an increase in the traveled distance actually means that the velocity increases. The average velocity increases up to $\sim 10\,\mathrm{m}/\mathrm{s}$ which is however still much lower than the velocity of a skyrmion driven by an equivalent direct current of $j_\mathrm{DC}\Theta_\mathrm{SH}=10/\sqrt{2}\,\mathrm{MA}/\mathrm{cm}^2$ corresponding to  $v_x=46.4\,\mathrm{m}/\mathrm{s}$ (cf. Supplementary Fig. 2 for the simulation).
The reason is that the skyrmions follow a soft ratchet mechanism that is distinct from typical classical ratchet mechanisms and not as efficient. In the following, we investigate this unique mechanism in more detail and show that it can be turned into a strict and more efficient ratchet mechanism by tweaking the racetrack geometry.\\
\\
\textbf{Comparison to classical ratchet mechanisms.}
Especially from the simulations at larger periods, it becomes apparent that the ratchet is soft: Backwards motion is allowed, even multiple steps at a time. The average velocity is essentially determined by the difference in paths for the forwards motion at the upper edge and the backwards motion at the lower racetrack edge. The average velocity is quite far from the ideal velocity of a hard ratchet that would be half the velocity of a skyrmion driven by a direct current with the magnitude $j_{x,\mathrm{max}}/\sqrt{2}$.

\begin{figure}[t!]
  \centering
  \includegraphics[width=\columnwidth]{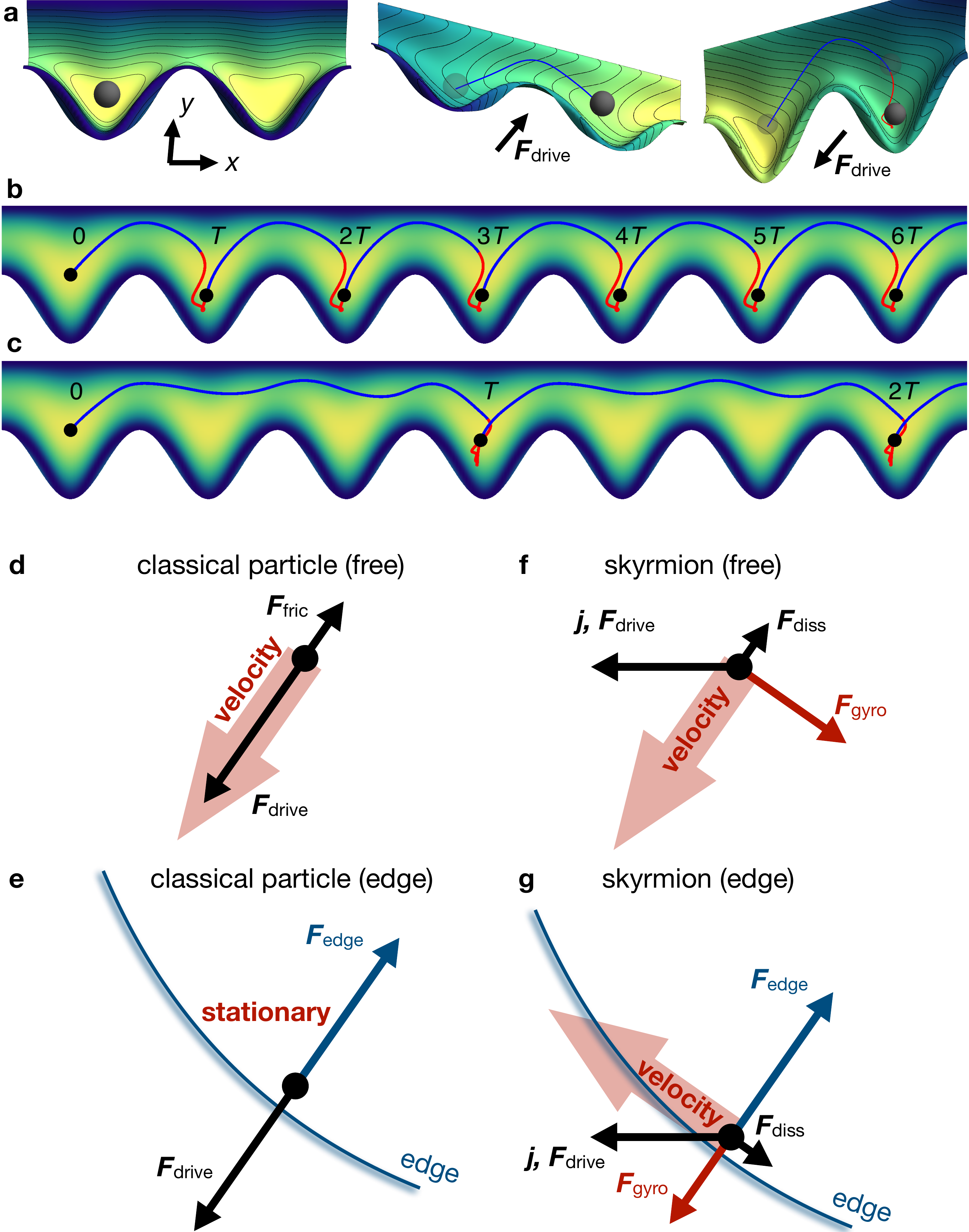}
  \caption{\textbf{Distinction from classical ratchet mechanisms.} \textbf{a-e} Motion of a classical particle in an asymmetric potential (color code) under the influence of a periodic driving force oriented at a polar angle of $50^\circ$. The distance between two pockets is $L=240\,\mathrm{m}$. \textbf{a}, Representation of the asymmetric potential including the effect of the driving force, $-\nabla U_\mathrm{drive}=\vec{F}_\mathrm{drive}$. The initial potential (left panel) is tilted (middle, right) so that the particle moves (blue and red trajectories) to the next cell. In \textbf{b}, for a period of the force $T=650\,\mathrm{s}$ the particle moves one cell per period. In \textbf{c}, for $T=1950\,\mathrm{s}$ the particle moves three cells per period. The colored lines indicate the trajectories for which blue corresponds to a positive force (along the positive $x$ and $y$ directions) and red corresponds to a negative force. \textbf{d}, The driving and friction forces for a free classical particle near the equilibrium position that is driven by a negative $\vec{F}_\mathrm{drive}$. \textbf{e} is similar to \textbf{d} but the classical particle is close to the confinement and feels an additional repulsive force $\vec{F}_\mathrm{edge}$. If this force compensates the driving force, the object can be stationary. \textbf{f-g}, Instead of a classical particle, a magnetic skyrmion is considered. As the main difference to the classical case, the gyroscopic force (red) is oriented differently for a free skyrmion and a skyrmion near the edge. As a consequence, the skyrmion close to the edge will drift along the edge and will not remain stationary.}
  \label{fig:classical}
\end{figure}

To further understand how and why the skyrmion ratchet mechanism differs from a strict ratchet, we investigate a simple toy-model of a strict ratchet in a geometry similar to ours
\begin{align}
m\ddot{\vec{r}}=\vec{F}_\mathrm{drive}-\xi\dot{\vec{r}}-\nabla U(\vec{r}).
\end{align}
Here $m=1\,\mathrm{kg}$ is the mass, $\vec{F}_\mathrm{drive}$ is a periodic driving force, $\xi=0.02\,\mathrm{kg}/\mathrm{s}$ is the friction constant,  and $U(\vec{r})$ is a potential that is qualitatively similar to the potential of a skyrmion in the asymmetric racetrack; cf. color code in Figs. \ref{fig:classical}a (analytical expression given in the Methods section).
The model is chosen such that it resembles the propagation mode of the skyrmion to a certain extend so that the differences between the two ratchet mechanisms become apparent. A physical system that would follow this equation of motion is a ball rolling in an asymmetrically shaped bowl, as shown in Fig. \ref{fig:classical}a. By tilting the bowl periodically back and forth, the ball can travel a net distance; cf. Fig. \ref{fig:classical}a-c.

Comparing this equation of motion with the Thiele equation \eqref{eq:thiele} describing a skyrmions, both equations have a friction or dissipation term antiparallel to the velocity $\propto -\dot{\vec{r}}$ and an interaction with the asymmetric potential $U(\vec{r})$. The main difference is that the classical model has a mass term $m\ddot{\vec{r}}$ while the Thiele equation is mass-less and instead has a gyroscopic term which is always perpendicular to the skyrmion velocity $-4\pi N_\mathrm{Sk}b\,\vec{e}_z\times\dot{\vec{r}}$. Furthermore,  both systems are driven by a periodic driving force. However, this force points along $\pm x$ for the skyrmion but is chosen to point at an angle for the classical model (cf. Fig. \ref{fig:classical}d).

In Fig. \ref{fig:classical}b a one-step propagation mode is shown for the toy model. The trajectory is similar to the skyrmionic case but for the the second half-period (red) the particle is not moving backwards along the lower edge but it is pushed towards that edge and remains almost stationary. 

When the period $T$ of the periodic force is increased, different propagation modes can be achieved. Exemplarily, in Fig. \ref{fig:classical}c a three-step mode is shown. Here, the particle moves three steps forwards in the first half-period (blue) and remains almost stationary in the second half period (red). This behavior is drastically different to the skyrmion propagation mode at high $T$, where the skyrmion moves backwards one or even multiple cells when the current is negative; cf. Fig. \ref{fig:modes}d-f.

The reason for this different behavior can be understood by investigating all individual forces for free classical particles and skyrmions, as well as both objects close to the edge. In the case of a free classical particle, the motion occurs parallel to the driving force (Fig. \ref{fig:classical}d). The only other force is the friction force that is always antiparallel to the velocity. When the particle approaches the edge, a new force becomes significant (Fig. \ref{fig:classical}e): This is $\vec{F}_\mathrm{edge}=-\nabla U(\vec{r})$. The particle moves closer and closer to the edge until $\vec{F}_\mathrm{edge}$ and $\vec{F}_\mathrm{drive}$ compensate each other. For the considered geometry, the particle cannot move to the cell to the left because it would have to overcome an energy barrier.

For the skyrmion in the free case (Fig. \ref{fig:classical}f), the motion is similar to a classical particle. However, as explained above, the velocity is not parallel to the driving force (that is here along $\pm \vec{x}$) but the topological charge leads to the emergence of a gyroscopic force that leads to a deflection towards the edge. When the skyrmion approaches the lower edge, an edge force $\vec{F}_\mathrm{edge}=-\nabla U(\vec{r})$ becomes significant just as for the classical particle. However, since this force affects the velocity, also the gyroscopic force changes its orientation (the reader is reminded that the gyroscopic force is always perpendicular to the velocity). As a consequence, when the skyrmion is pushed towards the edge, there is no total equilibrium of forces but a net force parallel to the edge once the skyrmion moves in a steady state close to the edge, as in indicated in Fig. \ref{fig:classical}g. The skyrmion will creep along the edge into the next pocket as was shown in the micromagnetic simulations in Fig. \ref{fig:modes}d-f. 

Motivated by the motion of the skyrmion parallel to both edges (deformed and undeformed), one can argue that the skyrmion moves in a topological edge channel. This mode of propulsion is also the reason why skyrmions \change{can} avoid defects, as has been simulated for example in Ref. \cite{sampaio2013nucleation}. \change{Note, that the interaction with skyrmions can be repulsive but also attractive \cite{fernandes2018universality}. For this reason, skyrmions can also be trapped by defects\cite{muller2015capturing,muller2016edge,hanneken2016pinning, castell2019accelerating}}. 

Furthermore, it is worth mentioning that these edge chanels are fundamentally different from those that have been discussed for frustrated magnets in Ref. \cite{leonov2017edge}. In that reference, the magnetization modulations close to the edge of the sample give rise to local minima in the potential $U(\vec{r})$. In the DMI mediated systems that we discuss here, such minima are absent and the distance of a steady-state skyrmion to the edge is even dependent on its velocity.\\
\\
\textbf{Constructing a strict skyrmion ratchet.}
The above presented ratchet behavior is interesting from a fundamental point of view because it shows new propagation dynamics distinct from classical strict ratchets. However, for typical spintronic applications it is unfavorable due to the low efficiency. The maximum velocity, determined in Fig. \ref{fig:modes}b is around $10\,\mathrm{m}/\mathrm{s}$ for $j_{x,\mathrm{max}}\Theta_\mathrm{SH}=10\,\mathrm{MA}/\mathrm{cm}^2$, while a skyrmion driven by equivalent direct currents $j_{x,\mathrm{DC}}\Theta_\mathrm{SH}=10/\sqrt{2}\,\mathrm{MA}/\mathrm{cm}^2$ moves at $46.4\,\mathrm{m}/\mathrm{s}$ (simulation presented in Supplementary Fig. 2). 
Therefore, now we present how the asymmetric racetrack geometry can be altered to achieve a strict skyrmion ratchet that allows for a faster propagation. 

\begin{figure}[t!]
  \centering
  \includegraphics[width=\columnwidth]{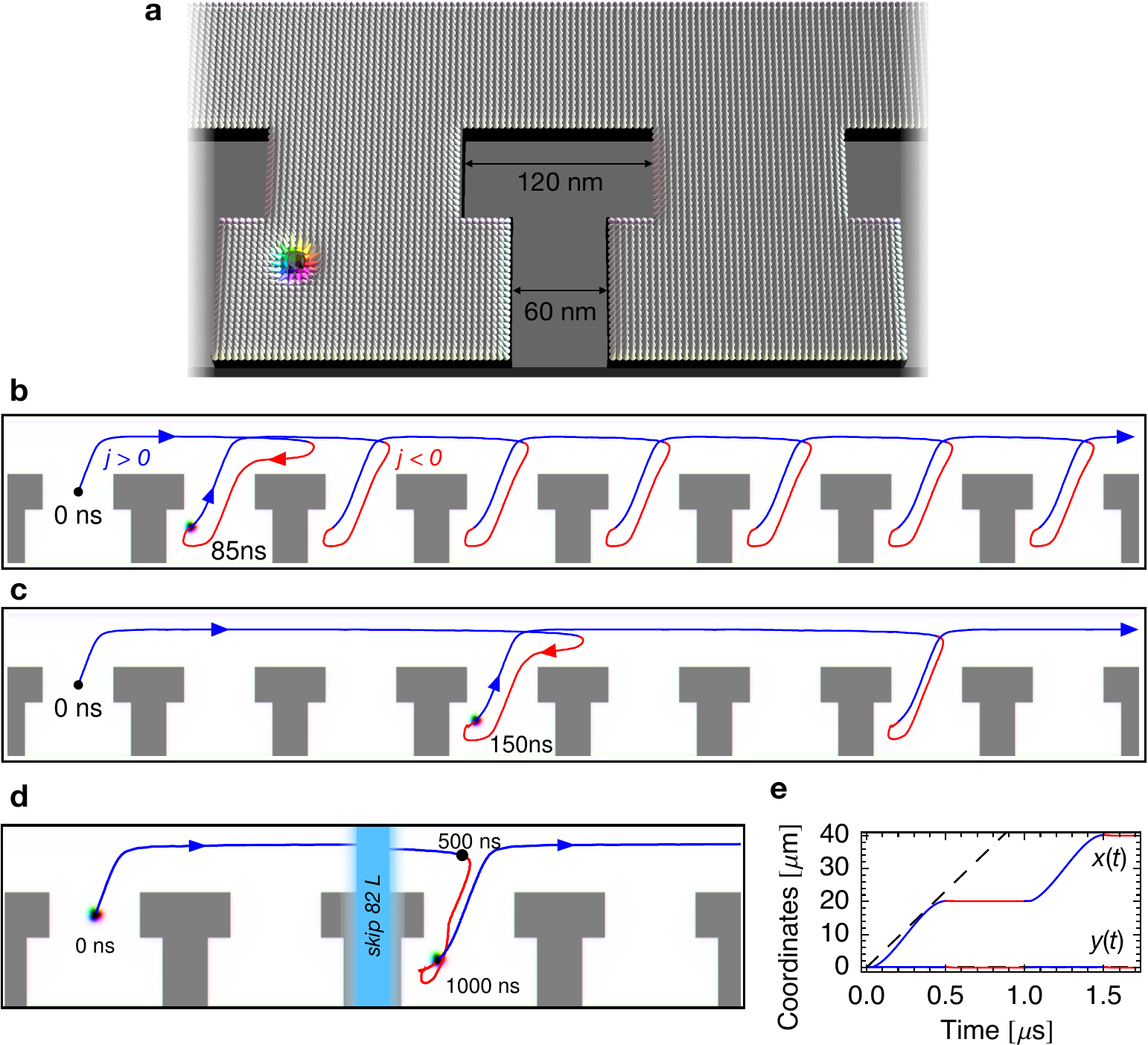}
  \caption{\textbf{Strict skyrmion ratchet.} \change{\textbf{a}, Skyrmion (colored arrows) in an asymmetric racetrack with modified pockets. \textbf{b}, One-step motion of a skyrmion under the influence of a periodic current with period $T=85\,\mathrm{ns}$ and amplitude $j_{x,\mathrm{max}}\Theta_\mathrm{SH}=3\,\frac{\mathrm{MA}}{\mathrm{cm}^2}$, similar to the one-step motion in Fig. \ref{fig:ratchet}a. \textbf{c}, Three-step motion for $T=150\,\mathrm{ns}$. Backwards propagation is suppressed making this propulsion mode more efficient than the two-step motion in Fig. \ref{fig:modes}d. \textbf{d}, Generalization of this concept of a strict ratchet motion. For $T=1\,\mu\mathrm{s}$ and $j_{x,\mathrm{max}}\Theta_\mathrm{SH}=10\,\frac{\mathrm{MA}}{\mathrm{cm}^2}$ the skyrmion moves 84 steps forwards and 0 steps backwards in a single period. The light blue area contains 82 pockets that are not shown for the sake of visibility. \textbf{e}, shows the corresponding $x$ and $y$ coordinates. The dashed line shows the propagation of a skyrmion under an equivalent direct current of $j_\mathrm{DC}\Theta_\mathrm{SH}=10/\sqrt{2}\,\frac{\mathrm{MA}}{\mathrm{cm}^2}$. Animated versions of the panels \textbf{b-d} are shown in the Supplementary Movies 8-10.}}
  \label{fig:strict}
\end{figure}

Ideally, in a strict ratchet 50\% of the energy is used for forwards translation and the remaining 50\% is \textit{not} used for backwards translation. To achieve this for the skyrmion, firstly the backwards motion must be suppressed, like in the toy model for the classical particle; cf. Fig. \ref{fig:classical}b,c. We achieve this by narrowing the pockets of the racetrack at the top; cf. Fig. \ref{fig:strict}a. While we have presented that skyrmions can creep up vertical walls -- different from the classical particle in the toy model -- there exists a limit. A skyrmion cannot creep along the horizontal edges at the top of the pockets and it gets stuck in the corners (Fig. \ref{fig:strict}a). 

\change{In Fig. \ref{fig:strict}b-d the one-step propagation mode and two multi-step propagation modes are shown. A direct comparison with the soft-ratchet mechanism in the initial geometry (Fig. \ref{fig:modes}c-f) shows that the strict ratchet mechanism can be much more efficient, since backwards propagation can only occur inside of a single pocket. In panel b, for $T=85\,\mathrm{ns}$, the skyrmion moves one cell size $L=240\,\mathrm{nm}$ per period. This type of motion is as efficient as the motion in the initial geometry, because there backwards propagation was only occurring within a single pocket. For higher $T$, the new geometry becomes more efficient. For example, in Fig. \ref{fig:strict}c, for $T=150\,\mathrm{ns}$, the skyrmion moves $3L$ per period while it moved only $2L$ in the initial geometry (cf. Fig. \ref{fig:modes}d; $3L$ forward and $1L$ backwards).}

\change{To optimize the efficiency of the mechanism}, the time for moving into and out of the pockets must be minimized in relation to the period. We achieve this by increasing the period duration $T$. 
In the simulation shown in Fig. \ref{fig:strict}\change{d} we have used $j_{x,\mathrm{max}}\Theta_\mathrm{SH}=10\,\mathrm{MA}/\mathrm{cm}^2$ and $T=1\,\mu\mathrm{s}$ \change{as an extreme case}. Within one period the skyrmion moves 84 cell sizes to the right and not a single cell to the left which corresponds to an average velocity of $20.2\,\mathrm{m}/\mathrm{s}$. This velocity is much faster than what was predicted in the skyrmion ratchets mentioned in the introduction \cite{ma2017reversible, wang2018efficient,moon2016skyrmion, chen2019skyrmion} and is also twice as fast as the skyrmion propulsion by the soft ratchet mechanism presented above. The determined average velocity is $43.5\%$ of the velocity of a skyrmion driven by an equivalent direct current $j_{x,\mathrm{DC}}\Theta_\mathrm{SH}=10/\sqrt{2}\,\mathrm{MA}/\mathrm{cm}^2$. This skyrmion moves at $46.4\,\mathrm{m}/\mathrm{s}$ (simulation presented in Supplementary Figure 2), which corresponds to the dashed line in Fig. \ref{fig:strict}\change{e}. In the first half period of the AC simulation, this velocity is reached. The established strict skyrmion ratchet is close to the efficiency maximum of $50\%$. It can be increased even further when $T$ is increased.\\
\\
\textbf{Discussion}\\
We have shown that the skyrmion Hall effect can be utilized to drive topologically non-trivial spin textures by AC currents, when the racetrack is not mirror-symmetric. The skyrmion Hall effect delivers the periodic oscillation of the $y$ coordinate; entering and exiting the pockets at the bottom edge of the racetrack at the beginning and the end of the negative current half-period. While the skyrmion moves freely along the racetrack for positive current densities, the backwards motion is decelerated when the current is negative. This leads to a net motion of one or even multiple racetrack pockets per period. This mechanism allows to precisely tune the skyrmion velocity because in a stable regime it is determined (besides by the propagation mode) by the current period and the cell size $\overline{v}=n\cdot L/T$, with $n=0,1,2,\dots$.

\change{The fundamental origin of the mechanism is the gyroscopic force that arises due to the topological charge, similar to what has been described in Refs. \cite{reichhardt2015magnus,reichhardt2020dynamics}, and the asymmetric environment that is here constituted by a modified racetrack geometry. Both ingredients together lead to a directed motion of skyrmions parallel to an alternating current. Since this motion relies on the topological properties of the spin textures, this ratchet mechanism can also be used to distinguish or even separate different types of spin objects. Topologically trivial skyrmioniums, bubbles or domain walls, for example, could be separated from skyrmions.}

 \change{Furthermore, due to the periodic racetrack geometry, a skyrmion-bit stream is `discretized' in the sense that once the current is turned off, a skyrmion will always relax to the middle of a pocket. Its position is much less susceptible to interactions with other skyrmions or random motion due to thermal fluctuation compared to conventional racetrack data storage designs. Also, this makes a controlled writing and reading of the bits more convenient.}

Especially for the propagation of multiple racetrack pockets per period it became apparent that the skyrmion follows a soft ratchet mechanism which means that backwards propagation is possible. This is in contrast to the typical strict ratchet that we have simulated using classical equations of motion. The main difference between the two systems is that the skyrmion's equation of motion, the Thiele equation, does not contain a mass term but a gyroscopic force that is always oriented perpendicular to the velocity of the skyrmion. For this reason, a skyrmion close to the edge will not experience a balance of forces but is pushed parallel to the edge and moves like in an edge channel.

This behavior is fundamentally interesting, because it reveals the special properties of skyrmions based on their non-trivial topology. However, due to the comparably slow propagation, it is technologically unfavorable for typical racetrack applications. Therefore, we have also shown that this behavior can be suppressed. By tweaking the racetrack pockets, the backwards propagation can be completely suppressed, so that almost half of the energy of the alternating current can be used for propagation. We have simulated average velocities of over $20\,\mathrm{m}/\mathrm{s}$ which is faster than what was reported for the skyrmion ratchets mentioned in the introduction \cite{ma2017reversible, wang2018efficient,moon2016skyrmion, chen2019skyrmion}.

In summary, we have shown that skyrmion-based racetrack storage devices can be operated by alternating currents via a ratchet mechanism. Using alternating currents instead of direct currents has proven to be advantageous in conventional electronics and further increases the technological significance of skyrmions for spintronics. \\
\\
\textbf{Methods}\\
\textbf{Micromagnetic simulations.}
The micromagnetic simulations have been conducted according to the Lifshitz-Gilbert equation (LLG)~\cite{landau1935theory,gilbert1955lagrangian,slonczewski1996current} 
\begin{align}
\dot{\vec{m}}_i=&-\gamma_e\vec{m}_i\times\vec{B}_{i,\mathrm{eff}}+\alpha\vec{m}_i\times\dot{\vec{m}_i}\label{eq:llg}\\
&+\gamma_e \epsilon\beta[(\vec{m}_i\times\vec{s})\times\vec{m}_i].\notag
\end{align}
This equation has been propagated using the micromagnetic simulation code mumax3 \cite{vansteenkiste2011mumax, vansteenkiste2014design}.

The LLG equation includes the gyromagnetic ratio of an electron $\gamma_e=\gamma/\mu_0=1.760\times 10^{11}\,\mathrm{T}^{-1}\mathrm{s}^{-1}$, the in-plane torque coefficient $\epsilon\beta=\frac{\hbar j\Theta_\mathrm{SH}}{2ed_zM_s}$, the Gilbert damping coefficient $\alpha=0.3$ and the effective magnetic field $\vec{B}_{i,\mathrm{eff}}=-\delta_{\vec{m}_i}F/M_s$. This field accounts for all magnetic interactions and it is determined by the free energy density $F$. For this energy we have included the exchange interaction, the interfacial Dzyaloshinskii-Moriya interaction~\cite{dzyaloshinsky1958thermodynamic,moriya1960anisotropic} (DMI), a uniaxial anisotropy and the effect of the demagnetization field.

The parameters for CoPt have been taken from Ref.~\onlinecite{sampaio2013nucleation}: exchange stiffness $J=15\,\mathrm{pJ}/\mathrm{m}$, DMI constant $D=3\,\mathrm{mJ}/\mathrm{m}^2$, uniaxial anisotropy $K_z=0.8\,\mathrm{MJ}/\mathrm{m}^3$, saturation magnetization $M_s=0.58\,\mathrm{MA}/\mathrm{m}$. Only for the simulation of the skyrmionium in Fig. 2b we have increased the DMI constant to $D=3.5\,\mathrm{mJ}/\mathrm{m}^2$, as was suggested in Ref.~\onlinecite{zhang2016control}. The spin Hall angle is $\Theta_\mathrm{SH}=0.3$.

In Fig. 2 of the main paper, the racetrack dimension are $1200\,\mathrm{nm}\times 240\,\mathrm{nm}\times 2\,\mathrm{nm}$. The added pockets are characterized by circles with a diameter of $120\,\mathrm{nm}$, as indicated in Fig. 1. In Fig. 3 and Fig. 4c,d,e the simulated length has been increased to $1920\,\mathrm{nm}$. In Fig. 4a,b the simulated length is only $240\,\mathrm{nm}$ to speed up the simulation. In Fig. 6\change{b,c} the simulated length is \change{$1920\,\mathrm{nm}$ as well, and in Fig. 6d} $960\,\mathrm{nm}$ and the sizes of the rectangular pockets are given in Fig. 6a. In all simulations, periodic boundary conditions have been considered along the track direction. The cell sizes are $2\,\mathrm{nm}\times 2\,\mathrm{nm}\times 2\,\mathrm{nm}$. \\
\\
\textbf{Toy model.}
For the simulation of the toy model we have propagated the equation of motion 
\begin{align}
m\ddot{\vec{r}}=\vec{F}_\mathrm{drive}-\xi\dot{\vec{r}}-\nabla U(\vec{r}).
\end{align}
The parameters are: mass $m=1\,\mathrm{kg}$, friction constant $\xi=0.02\,\mathrm{kg}/\mathrm{s}$, periodic driving force $\vec{F}_\mathrm{drive}=40\,\mathrm{mN}\sin(\frac{2\pi}{T})\vec{e}$ with $\vec{e}=\cos(50^\circ)\vec{e}_x+\sin(50^\circ)\vec{e}_y$ and $T=650\,\mathrm{s}$ for the 1-step mode (Fig. \ref{fig:classical}b) and $T=1950\,\mathrm{s}$ for the 3-step mode (Fig. \ref{fig:classical}c). The potential is
\begin{align*}
U(\vec{r})=U_0\,\exp\Bigg\{\frac{1}{4}\Big[1-\cos\Big(x\frac{2\pi}{L}\Big)\Big]-\frac{1}{4}\Big[3+\cos\Big(x\frac{2\pi}{L}\Big)\Big]\\
\times\cos
\Big\{\frac{\pi}{2}\Big[1-\cos\Big(x\frac{2\pi}{L}\Big)\Big]-y\frac{\pi}{L}\Big[3-\cos\Big(x\frac{2\pi}{L}\Big)\Big]\Big\}\Bigg\}.
\end{align*}
This potential has been constructed to account for the asymmetric geometry.
The amplitude is $U_0=1\,\mathrm{J}$ and the lattice constant is $L=240\,\mathrm{m}$. \\
\\
\textbf{Data availability}\\
Data that support the findings of this work are available from the corresponding author on request.\\
\\
\textbf{Code availability}\\
For the micromagnetic simulations we used the open-source code mumax3 available at https://mumax.github.io/.\\
The classical ratchet model has been propagated using a Mathematica code developed for this purpose.\\
\\
\textbf{Acknowledgements}\\
This work is supported by SFB TRR 227 of Deutsche Forschungsgemeinschaft (DFG). B.G. acknowledges valuable discussions with Jamal Berakdar and Yaroslav Pavlyukh about mechanical ratchets and with Stuart S. P. Parkin about ratchets in spintronics.\\
\\
\textbf{Author contributions}\\
B.G. initiated research, did the calculations and wrote the initial draft. I.M supervised the project. All authors discussed the results and contributed to the manuscript.\\
\\
\textbf{Supplementary information}\\
accompanies this paper at [insert link].\\
\\
\textbf{Competing interests}\\
The authors declare no competing interests.

%%\bibliography{short,MyLibrary}
%%\bibliographystyle{apsrev}

%\bibliography{short,MyLibrary}

\begin{thebibliography}{10}
\expandafter\ifx\csname url\endcsname\relax
  \def\url#1{\texttt{#1}}\fi
\expandafter\ifx\csname urlprefix\endcsname\relax\def\urlprefix{URL }\fi
\providecommand{\bibinfo}[2]{#2}
\providecommand{\eprint}[2][]{\url{#2}}

\bibitem{feynman1966feynman}
\bibinfo{author}{Feynman, R.~P.}, \bibinfo{author}{Leighton, R.~B.} \&
  \bibinfo{author}{Sands, M.}
\newblock \bibinfo{title}{The {F}eynman lectures on physics ch. 46 vol. 1}
  (\bibinfo{year}{1966}).

\bibitem{reimann2002brownian}
\bibinfo{author}{Reimann, P.}
\newblock \bibinfo{title}{Brownian motors: noisy transport far from
  equilibrium}.
\newblock \emph{\bibinfo{journal}{Physics Reports}}
  \textbf{\bibinfo{volume}{361}}, \bibinfo{pages}{57--265}
  (\bibinfo{year}{2002}).

\bibitem{hanggi2009artificial}
\bibinfo{author}{H{\"a}nggi, P.} \& \bibinfo{author}{Marchesoni, F.}
\newblock \bibinfo{title}{Artificial {B}rownian motors: Controlling transport
  on the nanoscale}.
\newblock \emph{\bibinfo{journal}{Reviews of Modern Physics}}
  \textbf{\bibinfo{volume}{81}}, \bibinfo{pages}{387} (\bibinfo{year}{2009}).

\bibitem{lavrijsen2013magnetic}
\bibinfo{author}{Lavrijsen, R.} \emph{et~al.}
\newblock \bibinfo{title}{Magnetic ratchet for three-dimensional spintronic
  memory and logic}.
\newblock \emph{\bibinfo{journal}{Nature}} \textbf{\bibinfo{volume}{493}},
  \bibinfo{pages}{647--650} (\bibinfo{year}{2013}).

\bibitem{flatte2008one}
\bibinfo{author}{Flatt{\'e}, M.~E.}
\newblock \bibinfo{title}{A one-way street for spin current}.
\newblock \emph{\bibinfo{journal}{Nature Physics}}
  \textbf{\bibinfo{volume}{4}}, \bibinfo{pages}{587--588}
  (\bibinfo{year}{2008}).

\bibitem{costache2010experimental}
\bibinfo{author}{Costache, M.~V.} \& \bibinfo{author}{Valenzuela, S.~O.}
\newblock \bibinfo{title}{Experimental spin ratchet}.
\newblock \emph{\bibinfo{journal}{Science}} \textbf{\bibinfo{volume}{330}},
  \bibinfo{pages}{1645--1648} (\bibinfo{year}{2010}).

\bibitem{himeno2008domain}
\bibinfo{author}{Himeno, A.}, \bibinfo{author}{Kondo, K.},
  \bibinfo{author}{Tanigawa, H.}, \bibinfo{author}{Kasai, S.} \&
  \bibinfo{author}{Ono, T.}
\newblock \bibinfo{title}{Domain wall ratchet effect in a magnetic wire with
  asymmetric notches}.
\newblock \emph{\bibinfo{journal}{Journal of Applied Physics}}
  \textbf{\bibinfo{volume}{103}}, \bibinfo{pages}{07E703}
  (\bibinfo{year}{2008}).

\bibitem{franken2012shift}
\bibinfo{author}{Franken, J.}, \bibinfo{author}{Swagten, H.} \&
  \bibinfo{author}{Koopmans, B.}
\newblock \bibinfo{title}{Shift registers based on magnetic domain wall
  ratchets with perpendicular anisotropy}.
\newblock \emph{\bibinfo{journal}{Nature Nanotechnology}}
  \textbf{\bibinfo{volume}{7}}, \bibinfo{pages}{499--503}
  (\bibinfo{year}{2012}).

\bibitem{piao2011ratchet}
\bibinfo{author}{Piao, H.-G.}, \bibinfo{author}{Choi, H.-C.},
  \bibinfo{author}{Shim, J.-H.}, \bibinfo{author}{Kim, D.-H.} \&
  \bibinfo{author}{You, C.-Y.}
\newblock \bibinfo{title}{Ratchet effect of the domain wall by asymmetric
  magnetostatic potentials}.
\newblock \emph{\bibinfo{journal}{Applied Physics Letters}}
  \textbf{\bibinfo{volume}{99}}, \bibinfo{pages}{192512}
  (\bibinfo{year}{2011}).

\bibitem{whyte2015diode}
\bibinfo{author}{Whyte, J.} \& \bibinfo{author}{Gregg, J.}
\newblock \bibinfo{title}{A diode for ferroelectric domain-wall motion}.
\newblock \emph{\bibinfo{journal}{Nature Communications}}
  \textbf{\bibinfo{volume}{6}}, \bibinfo{pages}{7361} (\bibinfo{year}{2015}).

\bibitem{bogdanov1989thermodynamically}
\bibinfo{author}{Bogdanov, A.} \& \bibinfo{author}{Yablonskii, D.}
\newblock \bibinfo{title}{Thermodynamically stable vortices in magnetically
  ordered crystals. the mixed state of magnets}.
\newblock \emph{\bibinfo{journal}{Zh. Eksp. Teor. Fiz}}
  \textbf{\bibinfo{volume}{95}}, \bibinfo{pages}{182} (\bibinfo{year}{1989}).

\bibitem{muhlbauer2009skyrmion}
\bibinfo{author}{M{\"u}hlbauer, S.} \emph{et~al.}
\newblock \bibinfo{title}{Skyrmion lattice in a chiral magnet}.
\newblock \emph{\bibinfo{journal}{Science}} \textbf{\bibinfo{volume}{323}},
  \bibinfo{pages}{915--919} (\bibinfo{year}{2009}).

\bibitem{yu2010real}
\bibinfo{author}{Yu, X.} \emph{et~al.}
\newblock \bibinfo{title}{Real-space observation of a two-dimensional skyrmion
  crystal}.
\newblock \emph{\bibinfo{journal}{Nature}} \textbf{\bibinfo{volume}{465}},
  \bibinfo{pages}{901--904} (\bibinfo{year}{2010}).

\bibitem{nagaosa2013topological}
\bibinfo{author}{Nagaosa, N.} \& \bibinfo{author}{Tokura, Y.}
\newblock \bibinfo{title}{Topological properties and dynamics of magnetic
  skyrmions}.
\newblock \emph{\bibinfo{journal}{Nature Nanotechnology}}
  \textbf{\bibinfo{volume}{8}}, \bibinfo{pages}{899--911}
  (\bibinfo{year}{2013}).

\bibitem{parkin2004shiftable}
\bibinfo{author}{Parkin, S. S.~P.}
\newblock \bibinfo{title}{Shiftable magnetic shift register and method of using
  the same} (\bibinfo{year}{2004}).
\newblock \bibinfo{note}{{US} Patent 6,834,005}.

\bibitem{parkin2008magnetic}
\bibinfo{author}{Parkin, S. S.~P.}, \bibinfo{author}{Hayashi, M.} \&
  \bibinfo{author}{Thomas, L.}
\newblock \bibinfo{title}{Magnetic domain-wall racetrack memory}.
\newblock \emph{\bibinfo{journal}{Science}} \textbf{\bibinfo{volume}{320}},
  \bibinfo{pages}{190--194} (\bibinfo{year}{2008}).

\bibitem{parkin2015memory}
\bibinfo{author}{Parkin, S. S.~P.} \& \bibinfo{author}{Yang, S.-H.}
\newblock \bibinfo{title}{Memory on the racetrack}.
\newblock \emph{\bibinfo{journal}{Nature Nanotechnology}}
  \textbf{\bibinfo{volume}{10}}, \bibinfo{pages}{195--198}
  (\bibinfo{year}{2015}).

\bibitem{sampaio2013nucleation}
\bibinfo{author}{Sampaio, J.}, \bibinfo{author}{Cros, V.},
  \bibinfo{author}{Rohart, S.}, \bibinfo{author}{Thiaville, A.} \&
  \bibinfo{author}{Fert, A.}
\newblock \bibinfo{title}{Nucleation, stability and current-induced motion of
  isolated magnetic skyrmions in nanostructures}.
\newblock \emph{\bibinfo{journal}{Nature Nanotechnology}}
  \textbf{\bibinfo{volume}{8}}, \bibinfo{pages}{839} (\bibinfo{year}{2013}).

\bibitem{fert2013skyrmions}
\bibinfo{author}{Fert, A.}, \bibinfo{author}{Cros, V.} \&
  \bibinfo{author}{Sampaio, J.}
\newblock \bibinfo{title}{Skyrmions on the track}.
\newblock \emph{\bibinfo{journal}{Nature Nanotechnol.}}
  \textbf{\bibinfo{volume}{8}}, \bibinfo{pages}{152--156}
  (\bibinfo{year}{2013}).

\bibitem{yu2017room}
\bibinfo{author}{Yu, G.} \emph{et~al.}
\newblock \bibinfo{title}{Room-temperature skyrmion shift device for memory
  application}.
\newblock \emph{\bibinfo{journal}{Nano Letters}} \textbf{\bibinfo{volume}{17}},
  \bibinfo{pages}{261--268} (\bibinfo{year}{2017}).

\bibitem{jonietz2010spin}
\bibinfo{author}{Jonietz, F.} \emph{et~al.}
\newblock \bibinfo{title}{Spin transfer torques in {M}n{S}i at ultralow current
  densities}.
\newblock \emph{\bibinfo{journal}{Science}} \textbf{\bibinfo{volume}{330}},
  \bibinfo{pages}{1648--1651} (\bibinfo{year}{2010}).

\bibitem{zang2011dynamics}
\bibinfo{author}{Zang, J.}, \bibinfo{author}{Mostovoy, M.},
  \bibinfo{author}{Han, J.~H.} \& \bibinfo{author}{Nagaosa, N.}
\newblock \bibinfo{title}{Dynamics of skyrmion crystals in metallic thin
  films}.
\newblock \emph{\bibinfo{journal}{Phys.\ Rev.\ Lett.}}
  \textbf{\bibinfo{volume}{107}}, \bibinfo{pages}{136804}
  (\bibinfo{year}{2011}).

\bibitem{iwasaki2013current}
\bibinfo{author}{Iwasaki, J.}, \bibinfo{author}{Mochizuki, M.} \&
  \bibinfo{author}{Nagaosa, N.}
\newblock \bibinfo{title}{Current-induced skyrmion dynamics in constricted
  geometries}.
\newblock \emph{\bibinfo{journal}{Nature Nanotechnology}}
  \textbf{\bibinfo{volume}{8}}, \bibinfo{pages}{742--747}
  (\bibinfo{year}{2013}).

\bibitem{jiang2017direct}
\bibinfo{author}{Jiang, W.} \emph{et~al.}
\newblock \bibinfo{title}{Direct observation of the skyrmion {H}all effect}.
\newblock \emph{\bibinfo{journal}{Nature Physics}}
  \textbf{\bibinfo{volume}{13}}, \bibinfo{pages}{162--169}
  (\bibinfo{year}{2017}).

\bibitem{litzius2017skyrmion}
\bibinfo{author}{Litzius, K.} \emph{et~al.}
\newblock \bibinfo{title}{Skyrmion {H}all effect revealed by direct
  time-resolved x-ray microscopy}.
\newblock \emph{\bibinfo{journal}{Nature Physics}}
  \textbf{\bibinfo{volume}{13}}, \bibinfo{pages}{170--175}
  (\bibinfo{year}{2017}).

\bibitem{tomasello2014strategy}
\bibinfo{author}{Tomasello, R.} \emph{et~al.}
\newblock \bibinfo{title}{A strategy for the design of skyrmion racetrack
  memories}.
\newblock \emph{\bibinfo{journal}{Scientific Reports}}
  \textbf{\bibinfo{volume}{4}}, \bibinfo{pages}{6784} (\bibinfo{year}{2014}).

\bibitem{zhang2015skyrmion}
\bibinfo{author}{Zhang, X.} \emph{et~al.}
\newblock \bibinfo{title}{Skyrmion-skyrmion and skyrmion-edge repulsions in
  skyrmion-based racetrack memory}.
\newblock \emph{\bibinfo{journal}{Scientific Reports}}
  \textbf{\bibinfo{volume}{5}}, \bibinfo{pages}{7643} (\bibinfo{year}{2015}).

\bibitem{gobel2018magnetic}
\bibinfo{author}{G{\"o}bel, B.}, \bibinfo{author}{Mook, A.},
  \bibinfo{author}{Henk, J.}, \bibinfo{author}{Mertig, I.} \&
  \bibinfo{author}{Tretiakov, O.~A.}
\newblock \bibinfo{title}{Magnetic bimerons as skyrmion analogues in in-plane
  magnets}.
\newblock \emph{\bibinfo{journal}{Phys.\ Rev.\ B}}
  \textbf{\bibinfo{volume}{99}}, \bibinfo{pages}{060407}
  (\bibinfo{year}{2019}).

\bibitem{gobel2020beyond}
\bibinfo{author}{G{\"o}bel, B.}, \bibinfo{author}{Mertig, I.} \&
  \bibinfo{author}{Tretiakov, O.~A.}
\newblock \bibinfo{title}{Beyond skyrmions: Review and perspectives of
  alternative magnetic quasiparticles}.
\newblock \emph{\bibinfo{journal}{arXiv preprint arXiv:2005.01390}}
  (\bibinfo{year}{2020}).

\bibitem{barker2016static}
\bibinfo{author}{Barker, J.} \& \bibinfo{author}{Tretiakov, O.~A.}
\newblock \bibinfo{title}{Static and dynamical properties of antiferromagnetic
  skyrmions in the presence of applied current and temperature}.
\newblock \emph{\bibinfo{journal}{Phys.\ Rev.\ Lett.}}
  \textbf{\bibinfo{volume}{116}}, \bibinfo{pages}{147203}
  (\bibinfo{year}{2016}).

\bibitem{zhang2016magnetic}
\bibinfo{author}{Zhang, X.}, \bibinfo{author}{Zhou, Y.} \&
  \bibinfo{author}{Ezawa, M.}
\newblock \bibinfo{title}{Magnetic bilayer-skyrmions without skyrmion {H}all
  effect}.
\newblock \emph{\bibinfo{journal}{Nature Communications}}
  \textbf{\bibinfo{volume}{7}}, \bibinfo{pages}{10293} (\bibinfo{year}{2016}).

\bibitem{zhang2016antiferromagnetic}
\bibinfo{author}{Zhang, X.}, \bibinfo{author}{Zhou, Y.} \&
  \bibinfo{author}{Ezawa, M.}
\newblock \bibinfo{title}{Antiferromagnetic skyrmion: stability, creation and
  manipulation}.
\newblock \emph{\bibinfo{journal}{Scientific Reports}}
  \textbf{\bibinfo{volume}{6}}, \bibinfo{pages}{24795} (\bibinfo{year}{2016}).

\bibitem{gobel2017afmskx}
\bibinfo{author}{G\"obel, B.}, \bibinfo{author}{Mook, A.},
  \bibinfo{author}{Henk, J.} \& \bibinfo{author}{Mertig, I.}
\newblock \bibinfo{title}{Antiferromagnetic skyrmion crystals: Generation,
  topological {H}all, and topological spin {H}all effect}.
\newblock \emph{\bibinfo{journal}{Phys.\ Rev.\ B}}
  \textbf{\bibinfo{volume}{96}}, \bibinfo{pages}{060406}
  (\bibinfo{year}{2017}).

\bibitem{legrand2020room}
\bibinfo{author}{Legrand, W.} \emph{et~al.}
\newblock \bibinfo{title}{Room-temperature stabilization of antiferromagnetic
  skyrmions in synthetic antiferromagnets}.
\newblock \emph{\bibinfo{journal}{Nature Materials}}
  \textbf{\bibinfo{volume}{19}}, \bibinfo{pages}{34--42}
  (\bibinfo{year}{2020}).

\bibitem{dohi2019formation}
\bibinfo{author}{Dohi, T.}, \bibinfo{author}{DuttaGupta, S.},
  \bibinfo{author}{Fukami, S.} \& \bibinfo{author}{Ohno, H.}
\newblock \bibinfo{title}{Formation and current-induced motion of synthetic
  antiferromagnetic skyrmion bubbles}.
\newblock \emph{\bibinfo{journal}{Nature Communications}}
  \textbf{\bibinfo{volume}{10}}, \bibinfo{pages}{5153} (\bibinfo{year}{2019}).

\bibitem{zhang2016control}
\bibinfo{author}{Zhang, X.} \emph{et~al.}
\newblock \bibinfo{title}{Control and manipulation of a magnetic skyrmionium in
  nanostructures}.
\newblock \emph{\bibinfo{journal}{Physical Review B}}
  \textbf{\bibinfo{volume}{94}}, \bibinfo{pages}{094420}
  (\bibinfo{year}{2016}).

\bibitem{goebel2019electrical}
\bibinfo{author}{G\"obel, B.}, \bibinfo{author}{Sch\"affer, A.},
  \bibinfo{author}{Berakdar, J.}, \bibinfo{author}{Mertig, I.} \&
  \bibinfo{author}{Parkin, S.}
\newblock \bibinfo{title}{Electrical writing, deleting, reading, and moving of
  magnetic skyrmioniums in a racetrack device}.
\newblock \emph{\bibinfo{journal}{Scientific Reports}}
  \textbf{\bibinfo{volume}{9}}, \bibinfo{pages}{12119} (\bibinfo{year}{2019}).

\bibitem{zhang2018real}
\bibinfo{author}{Zhang, S.}, \bibinfo{author}{Kronast, F.},
  \bibinfo{author}{van~der Laan, G.} \& \bibinfo{author}{Hesjedal, T.}
\newblock \bibinfo{title}{Real-space observation of skyrmionium in a
  ferromagnet-magnetic topological insulator heterostructure}.
\newblock \emph{\bibinfo{journal}{Nano Letters}} \textbf{\bibinfo{volume}{18}},
  \bibinfo{pages}{1057--1063} (\bibinfo{year}{2018}).

\bibitem{nayak2017magnetic}
\bibinfo{author}{Nayak, A.~K.} \emph{et~al.}
\newblock \bibinfo{title}{Magnetic antiskyrmions above room temperature in
  tetragonal {H}eusler materials}.
\newblock \emph{\bibinfo{journal}{Nature}} \textbf{\bibinfo{volume}{548}},
  \bibinfo{pages}{561} (\bibinfo{year}{2017}).

\bibitem{jena2020elliptical}
\bibinfo{author}{Jena, J.} \emph{et~al.}
\newblock \bibinfo{title}{Elliptical {B}loch skyrmion chiral twins in an
  antiskyrmion system}.
\newblock \emph{\bibinfo{journal}{Nature Communications}}
  \textbf{\bibinfo{volume}{11}}, \bibinfo{pages}{1115} (\bibinfo{year}{2020}).

\bibitem{kharkov2017bound}
\bibinfo{author}{Kharkov, Y.}, \bibinfo{author}{Sushkov, O.} \&
  \bibinfo{author}{Mostovoy, M.}
\newblock \bibinfo{title}{Bound states of skyrmions and merons near the
  {L}ifshitz point}.
\newblock \emph{\bibinfo{journal}{Phys.\ Rev.\ Lett.}}
  \textbf{\bibinfo{volume}{119}}, \bibinfo{pages}{207201}
  (\bibinfo{year}{2017}).

\bibitem{gao2019creation}
\bibinfo{author}{Gao, N.} \emph{et~al.}
\newblock \bibinfo{title}{Creation and annihilation of topological meron pairs
  in in-plane magnetized films}.
\newblock \emph{\bibinfo{journal}{Nature Communications}}
  \textbf{\bibinfo{volume}{10}}, \bibinfo{pages}{5603} (\bibinfo{year}{2019}).

\bibitem{wang2018efficient}
\bibinfo{author}{Wang, X.} \emph{et~al.}
\newblock \bibinfo{title}{Efficient skyrmion transport mediated by a voltage
  controlled magnetic anisotropy gradient}.
\newblock \emph{\bibinfo{journal}{Nanoscale}} \textbf{\bibinfo{volume}{10}},
  \bibinfo{pages}{733--740} (\bibinfo{year}{2018}).

\bibitem{ma2017reversible}
\bibinfo{author}{Ma, X.}, \bibinfo{author}{Reichhardt, C.~O.} \&
  \bibinfo{author}{Reichhardt, C.}
\newblock \bibinfo{title}{Reversible vector ratchets for skyrmion systems}.
\newblock \emph{\bibinfo{journal}{Physical Review B}}
  \textbf{\bibinfo{volume}{95}}, \bibinfo{pages}{104401}
  (\bibinfo{year}{2017}).

\bibitem{moon2016skyrmion}
\bibinfo{author}{Moon, K.-W.} \emph{et~al.}
\newblock \bibinfo{title}{Skyrmion motion driven by oscillating magnetic
  field}.
\newblock \emph{\bibinfo{journal}{Scientific Reports}}
  \textbf{\bibinfo{volume}{6}}, \bibinfo{pages}{20360} (\bibinfo{year}{2016}).

\bibitem{chen2019skyrmion}
\bibinfo{author}{Chen, W.}, \bibinfo{author}{Liu, L.}, \bibinfo{author}{Ji, Y.}
  \& \bibinfo{author}{Zheng, Y.}
\newblock \bibinfo{title}{Skyrmion ratchet effect driven by a biharmonic
  force}.
\newblock \emph{\bibinfo{journal}{Physical Review B}}
  \textbf{\bibinfo{volume}{99}}, \bibinfo{pages}{064431}
  (\bibinfo{year}{2019}).

\bibitem{mochizuki2014thermally}
\bibinfo{author}{Mochizuki, M.} \emph{et~al.}
\newblock \bibinfo{title}{Thermally driven ratchet motion of a skyrmion
  microcrystal and topological magnon hall effect}.
\newblock \emph{\bibinfo{journal}{Nature Materials}}
  \textbf{\bibinfo{volume}{13}}, \bibinfo{pages}{241--246}
  (\bibinfo{year}{2014}).

\bibitem{zhao2020ferromagnetic}
\bibinfo{author}{Zhao, L.}, \bibinfo{author}{Liang, X.}, \bibinfo{author}{Xia,
  J.}, \bibinfo{author}{Zhao, G.} \& \bibinfo{author}{Zhou, Y.}
\newblock \bibinfo{title}{A ferromagnetic skyrmion-based diode with a
  voltage-controlled potential barrier}.
\newblock \emph{\bibinfo{journal}{Nanoscale}} \textbf{\bibinfo{volume}{12}},
  \bibinfo{pages}{9507--9516} (\bibinfo{year}{2020}).

\bibitem{reichhardt2015magnus}
\bibinfo{author}{Reichhardt, C.}, \bibinfo{author}{Ray, D.} \&
  \bibinfo{author}{Reichhardt, C.~O.}
\newblock \bibinfo{title}{Magnus-induced ratchet effects for skyrmions
  interacting with asymmetric substrates}.
\newblock \emph{\bibinfo{journal}{New Journal of Physics}}
  \textbf{\bibinfo{volume}{17}}, \bibinfo{pages}{073034}
  (\bibinfo{year}{2015}).

\bibitem{reichhardt2020dynamics}
\bibinfo{author}{Reichhardt, C.} \& \bibinfo{author}{Reichhardt, C.}
\newblock \bibinfo{title}{Dynamics of magnus-dominated particle clusters,
  collisions, pinning, and ratchets}.
\newblock \emph{\bibinfo{journal}{Physical Review E}}
  \textbf{\bibinfo{volume}{101}}, \bibinfo{pages}{062602}
  (\bibinfo{year}{2020}).

\bibitem{landau1935theory}
\bibinfo{author}{Landau, L.~D.} \& \bibinfo{author}{Lifshitz, E.}
\newblock \bibinfo{title}{On the theory of the dispersion of magnetic
  permeability in ferromagnetic bodies}.
\newblock \emph{\bibinfo{journal}{Phys. Z. Sowjetunion}}
  \textbf{\bibinfo{volume}{8}}, \bibinfo{pages}{101--114}
  (\bibinfo{year}{1935}).

\bibitem{gilbert1955lagrangian}
\bibinfo{author}{Gilbert, T.}
\newblock \bibinfo{title}{A lagrangian formulation of the gyromagnetic equation
  of the magnetization field}.
\newblock \emph{\bibinfo{journal}{Physical Review}}
  \textbf{\bibinfo{volume}{100}}, \bibinfo{pages}{1243} (\bibinfo{year}{1955}).

\bibitem{slonczewski1996current}
\bibinfo{author}{Slonczewski, J.~C.}
\newblock \bibinfo{title}{Current-driven excitation of magnetic multilayers}.
\newblock \emph{\bibinfo{journal}{Journal of Magnetism and Magnetic Materials}}
  \textbf{\bibinfo{volume}{159}}, \bibinfo{pages}{L1--L7}
  (\bibinfo{year}{1996}).

\bibitem{vansteenkiste2011mumax}
\bibinfo{author}{Vansteenkiste, A.} \& \bibinfo{author}{Van~de Wiele, B.}
\newblock \bibinfo{title}{Mumax: a new high-performance micromagnetic
  simulation tool}.
\newblock \emph{\bibinfo{journal}{J. Magn.\ Magn.\ Mater.}}
  \textbf{\bibinfo{volume}{323}}, \bibinfo{pages}{2585--2591}
  (\bibinfo{year}{2011}).

\bibitem{vansteenkiste2014design}
\bibinfo{author}{Vansteenkiste, A.} \emph{et~al.}
\newblock \bibinfo{title}{The design and verification of mumax3}.
\newblock \emph{\bibinfo{journal}{AIP Adv.}} \textbf{\bibinfo{volume}{4}},
  \bibinfo{pages}{107133} (\bibinfo{year}{2014}).

\bibitem{thiele1973steady}
\bibinfo{author}{Thiele, A.}
\newblock \bibinfo{title}{Steady-state motion of magnetic domains}.
\newblock \emph{\bibinfo{journal}{Phys.\ Rev.\ Lett.}}
  \textbf{\bibinfo{volume}{30}}, \bibinfo{pages}{230} (\bibinfo{year}{1973}).

\bibitem{gobel2018overcoming}
\bibinfo{author}{G{\"o}bel, B.}, \bibinfo{author}{Mook, A.},
  \bibinfo{author}{Henk, J.} \& \bibinfo{author}{Mertig, I.}
\newblock \bibinfo{title}{Overcoming the speed limit in skyrmion racetrack
  devices by suppressing the skyrmion {H}all effect}.
\newblock \emph{\bibinfo{journal}{Phys.\ Rev.\ B}}
  \textbf{\bibinfo{volume}{99}}, \bibinfo{pages}{020405}
  (\bibinfo{year}{2019}).

\bibitem{fernandes2018universality}
\bibinfo{author}{Fernandes, I.~L.}, \bibinfo{author}{Bouaziz, J.},
  \bibinfo{author}{Bl{\"u}gel, S.} \& \bibinfo{author}{Lounis, S.}
\newblock \bibinfo{title}{Universality of defect-skyrmion interaction
  profiles}.
\newblock \emph{\bibinfo{journal}{Nature communications}}
  \textbf{\bibinfo{volume}{9}}, \bibinfo{pages}{4395} (\bibinfo{year}{2018}).

\bibitem{muller2015capturing}
\bibinfo{author}{M{\"u}ller, J.} \& \bibinfo{author}{Rosch, A.}
\newblock \bibinfo{title}{Capturing of a magnetic skyrmion with a hole}.
\newblock \emph{\bibinfo{journal}{Physical Review B}}
  \textbf{\bibinfo{volume}{91}}, \bibinfo{pages}{054410}
  (\bibinfo{year}{2015}).

\bibitem{muller2016edge}
\bibinfo{author}{M{\"u}ller, J.}, \bibinfo{author}{Rosch, A.} \&
  \bibinfo{author}{Garst, M.}
\newblock \bibinfo{title}{Edge instabilities and skyrmion creation in magnetic
  layers}.
\newblock \emph{\bibinfo{journal}{New Journal of Physics}}
  \textbf{\bibinfo{volume}{18}}, \bibinfo{pages}{065006}
  (\bibinfo{year}{2016}).

\bibitem{hanneken2016pinning}
\bibinfo{author}{Hanneken, C.}, \bibinfo{author}{Kubetzka, A.},
  \bibinfo{author}{Von~Bergmann, K.} \& \bibinfo{author}{Wiesendanger, R.}
\newblock \bibinfo{title}{Pinning and movement of individual nanoscale magnetic
  skyrmions via defects}.
\newblock \emph{\bibinfo{journal}{New Journal of Physics}}
  \textbf{\bibinfo{volume}{18}}, \bibinfo{pages}{055009}
  (\bibinfo{year}{2016}).

\bibitem{castell2019accelerating}
\bibinfo{author}{Castell-Queralt, J.}, \bibinfo{author}{Gonz{\'a}lez-G{\'o}mez,
  L.}, \bibinfo{author}{Del-Valle, N.}, \bibinfo{author}{Sanchez, A.} \&
  \bibinfo{author}{Navau, C.}
\newblock \bibinfo{title}{Accelerating, guiding, and compressing skyrmions by
  defect rails}.
\newblock \emph{\bibinfo{journal}{Nanoscale}} \textbf{\bibinfo{volume}{11}},
  \bibinfo{pages}{12589--12594} (\bibinfo{year}{2019}).

\bibitem{chen2017skyrmion}
\bibinfo{author}{Chen, X.} \emph{et~al.}
\newblock \bibinfo{title}{Skyrmion dynamics in width-varying nanotracks and
  implications for skyrmionic applications}.
\newblock \emph{\bibinfo{journal}{Applied Physics Letters}}
  \textbf{\bibinfo{volume}{111}}, \bibinfo{pages}{202406}
  (\bibinfo{year}{2017}).

\bibitem{iwasaki2014colossal}
\bibinfo{author}{Iwasaki, J.}, \bibinfo{author}{Koshibae, W.} \&
  \bibinfo{author}{Nagaosa, N.}
\newblock \bibinfo{title}{Colossal spin transfer torque effect on skyrmion
  along the edge}.
\newblock \emph{\bibinfo{journal}{Nano Letters}} \textbf{\bibinfo{volume}{14}},
  \bibinfo{pages}{4432--4437} (\bibinfo{year}{2014}).

\bibitem{navau2016analytical}
\bibinfo{author}{Navau, C.}, \bibinfo{author}{Del-Valle, N.} \&
  \bibinfo{author}{Sanchez, A.}
\newblock \bibinfo{title}{Analytical trajectories of skyrmions in confined
  geometries: Skyrmionic racetracks and nano-oscillators}.
\newblock \emph{\bibinfo{journal}{Physical Review B}}
  \textbf{\bibinfo{volume}{94}}, \bibinfo{pages}{184104}
  (\bibinfo{year}{2016}).

\bibitem{song2017skyrmion}
\bibinfo{author}{Song, C.} \emph{et~al.}
\newblock \bibinfo{title}{Skyrmion-based multi-channel racetrack}.
\newblock \emph{\bibinfo{journal}{Applied Physics Letters}}
  \textbf{\bibinfo{volume}{111}}, \bibinfo{pages}{192413}
  (\bibinfo{year}{2017}).

\bibitem{muller2017magnetic}
\bibinfo{author}{M{\"u}ller, J.}
\newblock \bibinfo{title}{Magnetic skyrmions on a two-lane racetrack}.
\newblock \emph{\bibinfo{journal}{New Journal of Physics}}
  \textbf{\bibinfo{volume}{19}}, \bibinfo{pages}{025002}
  (\bibinfo{year}{2017}).

\bibitem{leonov2017edge}
\bibinfo{author}{Leonov, A.} \& \bibinfo{author}{Mostovoy, M.}
\newblock \bibinfo{title}{Edge states and skyrmion dynamics in nanostripes of
  frustrated magnets}.
\newblock \emph{\bibinfo{journal}{Nature Communications}}
  \textbf{\bibinfo{volume}{8}}, \bibinfo{pages}{14394} (\bibinfo{year}{2017}).

\bibitem{dzyaloshinsky1958thermodynamic}
\bibinfo{author}{Dzyaloshinsky, I.}
\newblock \bibinfo{title}{A thermodynamic theory of “weak” ferromagnetism
  of antiferromagnetics}.
\newblock \emph{\bibinfo{journal}{J. Phys.\ Chem.\ Sol.}}
  \textbf{\bibinfo{volume}{4}}, \bibinfo{pages}{241--255}
  (\bibinfo{year}{1958}).

\bibitem{moriya1960anisotropic}
\bibinfo{author}{Moriya, T.}
\newblock \bibinfo{title}{Anisotropic superexchange interaction and weak
  ferromagnetism}.
\newblock \emph{\bibinfo{journal}{Phys.\ Rev.}} \textbf{\bibinfo{volume}{120}},
  \bibinfo{pages}{91} (\bibinfo{year}{1960}).

\end{thebibliography}
%\bibliographystyle{naturemag}

\end{document}